%Paper: hep-th/9311117
%From: dij@s-a.amtp.liv.ac.uk
%Date: Fri, 19 Nov 1993 12:32:30 +0000 (GMT)

\input phyzzx.tex

\def\apny{Ann.\ Phys.\ (New York)\ }
\def\cmp{Comm.\ Math.\ Phys.\ }

\def\npb{{Nucl.\ Phys.\ }{\bf B}}

\def\plb{{Phys.\ Lett.\ }{\bf B}}

\def\prd{{Phys.\ Rev.\ }{\bf D}}
\def\prl{Phys.\ Rev.\ Lett.\ }

\def \b{\beta}
\def \z{\zeta}
\def \m{\mu}
\def \a{\alpha}
\def \s{\sigma}
\def \e{\epsilon}
\def \P{\Phi}
\def \bP{\bar \Phi}
\def \bp{\bar p}
\def \bq{\bar q}
\def \br{\bar r}
\def \bs{\bar s}

\def \bv{\bar v}
\def \d{\delta}
\def \G{\Gamma}
\def \br{\bar r}
\def \t{\theta}
\def \bt{\bar \theta}
\font \bigbf=cmbx10 scaled \magstep2
{\nopagenumbers
\line {\hfil LTH 287}
\line {\hfil September 1992}
\vskip .5in
\centerline{\bigbf SIX-LOOP DIVERGENCES}
\centerline{\bigbf IN THE SUPERSYMMETRIC K\"AHLER SIGMA MODEL}
\vskip 1in
\centerline {\bf I. Jack, D. R. T. Jones and J. Panvel}
\vskip 5pt
\centerline {\it DAMTP, University of Liverpool, Liverpool L69 3BX, UK}
\vskip 1in
\centerline {\bf Abstract}
\vskip 5pt
The two-dimensional
supersymmetric $\s$-model on a K\"ahler manifold has a non-vanishing
$\b$-function at four loops, but the $\b$-function at five loops can be made
to vanish by a specific choice of renormalisation scheme. We investigate
whether this phenomenon persists at six loops, and conclude that it does not;
there is a non-vanishing six-loop $\b$-function irrespective of
 renormalisation scheme ambiguities.

\vfill
\eject}
\pageno=1
\tolerance=500
\line {\bigbf 1. Introduction \hfil}
\vskip 10pt
Two-dimensional non-linear $\s$-models have been the object of intense study,
in
recent years largely
because of their
relationship with string theory. A string propagating on a manifold $M$ is
described by a two-dimensional non-linear $\s$-model with $M$ as
target manifold. Interest has naturally focussed on supersymmetric $\s$-models
since the corresponding superstrings have desirable theoretical and
phenomenological properties, such as finiteness and anomaly cancellation.
Moreover, to obtain a
realistic theory, the ten-dimensional space on which the superstring
propagates must be compactified--in other words the string vacuum state must
be a manifold of the form $M_4\times K_6$, where $M_4$ is a maximally symmetric
four-dimensional space and $K_6$ is a six-dimensional manifold representing
internal compactified degrees of freedom. The requirement that the
four-dimensional manifold retains $N=1$ supersymmetry, which provides a
possible resolution of the ``naturalness'' problem\Ref\Witt{E. Witten, \npb188
(1981) 513}, then implies that $M_4$ is
Minkowski space and that $K_6$ is a Ricci-flat K\"ahler manifold\Ref\cand{P.
Candelas, G. T. Horowitz, A. Strominger and E. Witten, \npb258 (1985) 46}.
 We are thus led
to consider supersymmetric $\s$-models with a K\"ahler manifold as target
space; such theories in fact possess $N=2$ supersymmetry\Ref\zum{B. Zumino,
\plb87 (1979) 203}. We need to
determine the conditions for a manifold $M_4\times K_6$ to be a viable string
vacuum state; in fact this requires the corresponding $\s$-model to be
conformally invariant, which in turn implies that the renormalisation group
$\b$-functions for the K\"ahler metric should vanish\Ref\conf{C. Lovelace,
\plb136 (1984) 75; E. S. Fradkin and A. A. Tseytlin, \plb158 (1985) 316;
\plb160 (1985) 69; \npb261 (1985) 1; C. G. Callan, D. Friedan, E. Martinec
and M. J. Perry, \npb262 (1985) 593; A. Sen, \prd32 (1985) 210; \prl55 (1985)
 1846; \npb278 (1986) 289}
 (up to a diffeomorphism\Ref\gms{A. A. Tseytlin, \plb178 (1986) 34; G. M.
Shore, \npb286 (1987) 349}).
It was initially believed that the $\b$-function for a Ricci-flat
supersymmetric K\"ahler $\s$-model automatically vanished to all orders.
\Ref\agp{L. Alvarez-Gaum\'e and P. Ginsparg, \cmp102 (1985) 311}
However, Grisaru, van de Ven and Zanon
\Ref\gvzo{M. T. Grisaru, A. E. M. van de Ven and D. Zanon, \plb173 (1986)
 423}
\Ref\gvz{M. T. Grisaru, A. E. M. van de Ven and D. Zanon, \npb277 (1986) 388}
found a non-zero
contribution to the $\b$-function for the
supersymmetric K\"ahler $\s$-model at the four-loop level, which
did not vanish in the Ricci-flat case. In other words the natural metric on a
Ricci-flat K\"ahler manifold, (i.e. the one which {\it is} Ricci-flat),
does not satisfy the conformal invariance condition. Nevertheless Ricci-flat
K\"ahler manifolds may still be of phenomenological interest, since a metric
can be constructed on such manifolds (by adding non-local corrections to the
standard metric) which does satisfy the conformal invariance condition
\Ref\nem{D. Nemeschsansky and A. Sen, \plb178 (1986) 365}.
It was subsequently shown\Ref\gkz{M. T. Grisaru, D. I. Kazakov and D. Zanon,
\npb287 (1987) 189} that, remarkably, the five-loop divergence in the
K\"ahler $\s$-model could be removed by a {\it local} finite
redefinition of the metric in terms of covariant quantities,
 equivalent to a change of renormalisation scheme.
This result appears rather miraculous, and it is natural to ask whether it is
an isolated occurrence; might it be that there exists a scheme in which there
are no contributions to the $\b$-function beyond four loops? With
this motivation, we have carried out a partial computation of the six-loop
contribution to the $\b$-function for the K\"ahler $\s$-model. The terms we
have calculated are sufficient to determine that the six-loop $\b$-function
cannot be cancelled by a local covariant
field redefinition of the metric; there is no
renormalisation scheme in which the $\b$-function vanishes at six loops.
\vskip 1in
\line {\bigbf 2. Perturbative calculations for the K\"ahler sigma-model
\hfil}
\vskip 10pt
Our calculational methods are based on the work of Refs.~\gvz, ~\gkz. Firstly
we
describe the rudiments of K\"ahler geometry. A K\"ahler manifold is
a complex manifold with
a covariantly constant hermitian
almost complex structure, i.e. there is a tensor
$J_i{}^j$ satisfying
$$\eqalignno{
    J_i{}^kJ_k{}^j&=-\d_i{}^j,\cr
     J_i{}^kg_{kj}&=-J_j{}^kg_{ki},&(2.1)\cr
      \nabla_iJ_j{}^k&=0. \cr}
$$
We can then choose a local complex co-ordinate system $\P^p$, $\bP^{\bp}$, in
which the metric takes the form
$$\eqalignno{
g_{p\bq}&={\partial\over{\partial\P^p}}{\partial\over{\partial\bP^{\bq}}}
          K(\P,\bP),\cr
      \qquad g_{pq}&=g_{\bp\bq}=0 &(2.2)\cr}
$$
for some $K(\P,\bP)$ which is referred to as the K\"ahler potential.
Introducing the notation
$$
K_p\equiv{\partial K\over{\partial\P^p}},\qquad
K_{\bp}\equiv{\partial K\over{\partial\bP^{\bp}}},\eqno(2.3)
$$
the only non-vanishing Christoffel symbols are
$$
\G^p{}_{qr}=g^{p\bp}K_{\bp qr},\qquad \G^{\bp}{}_{\bq\br}=g^{\bp p}
K_{p\bq\br} \eqno(2.4)
$$
and the Riemann tensor is given by
$$
R_{p\bp q\bq}=K_{p\bp q\bq}-g^{r\br}K_{\br pq}K_{r\bp\bq}. \eqno(2.5)
$$
As we mentioned earlier, the $N=1$ supersymmetric $\s$-model defined on a
K\"ahler manifold in fact automatically possesses $N=2$ supersymmetry\refmark
\zum.
It can be expressed in terms of $N=2$ chiral and anti-chiral superfields
$\P^p(x,\t,\bt)$ and $\bP^{\bp}(x,\t,\bt)$ as
$$
S=\int d^2xd^2\t d^2\bt K(\P,\bP). \eqno(2.6)
$$
The chirality condition is
$$
\bar D_{\a}\P^p=D_{\a}\bP^{\bp}=0, \eqno(2.7)
$$
where the superspace covariant derivatives $D$, $\bar D$ are defined by
$$
D_{\a}={\partial\over{\partial\t_{\a}}}+{1\over2}i\bt^{\b}\partial_{\a\b},
\qquad\bar D_{\b}=(D_{\b})^*, \quad{\rm where}\quad
\partial_{\a\b}=\partial_{\m}\s^{\m}{}_{\a\b}.\eqno(2.8)
$$
(For notation and conventions see ``Superspace''\Ref\gates{S. J. Gates,
M. T. Grisaru, M. Ro\v cek and W. Siegel, Superspace (Benjamin-Cummings,
Reading,
Massachusetts, 1983)}.)
To perform perturbative calculations, we use the standard background field
method, expanding around a background $\P_0$, $\bP_0$ using a linear
quantum-background splitting
$$
\P\rightarrow\P_0+\P, \qquad \bP\rightarrow\bP_0+\bP. \eqno(2.9)
$$
The resulting expansion is then not manifestly covariant since the quantum
fields $\P^p$, $\bP^{\bp}$ are not vectors. This is in contrast to the normal
co-ordinate method\Ref\agfm{L. Alvarez-Gaum\'e, D. Z. Freedman and S. Mukhi,
\apny134 (1981) 85}\Ref\muk{S. Mukhi, \npb264 (1986) 640}
 usually adopted for the quantisation of the non-linear
$\s$-model; there the quantum field {\it is} a vector and consequently the
coefficients of the expansion are functions of the Riemann tensor and its
derivatives. This technique cannot be applied here due to the chirality
constraints Eq. (2.7). In any case, any consequent loss of elegance is
amply compensated by the many simplifications afforded by $N=2$ perturbation
theory. Moreover, the action Eq. (2.6) and its expansion are very compact and
it is relatively easy to recover a covariant expression at the end of the
calculation.
The expansion of the K\"ahler potential is then
$$\eqalignno{
K(\P_0+\P,\bP_0+\bP)-K(\P_0,\bP_0)&=K_p\P^p+K_{\bp}\bP^{\bp}\cr
&\quad +K_{p\bq}\P^p
\bP^{\bq}+{1\over2}K_{pq}\P^p\P^q+{1\over2}K_{\bp\bq}\bP^{\bp}\bP^{\bq}
+\ldots&(2.10)\cr}
$$
where we omit the dependence of $K$ on $\P_0$ and $\bP_0$ on the right-hand
side. The first quadratic term in Eq. (2.10) can be shown\refmark\gvz\
to give rise to an
effective propagator
$$
<\P^p(z)\bP^{\bq}(z')>=-g^{p\bq}\partial^{-2}D^2\d(z-z')\bar D^2 \eqno(2.11)
$$
where $z=(x,\t)$. The
remaining terms in the expansion then supply the vertices used to construct
Feynman diagrams. After standard $D$-algebra manipulations, the diagrams can
be written in momentum space form and hence evaluated. We use dimensional
regularisation so that we work in $d$ dimensions and divergences appear as
poles in $\e=2-d$. We construct counterterm diagrams on a
diagram-by-diagram basis, i.e. at each succeeding loop order, for each
diagram we write down counterterm diagrams corresponding to the subdivergences
of the original diagram. The remaining overall divergence is then cancelled
loop by loop by adding corrections to the K\"ahler potential, writing
$$
K\rightarrow K^B=K+\sum_{L=1}^{\infty}\sum_{n=1}^L{1\over{(4\pi)^L}}{K^{(n,L)}
\over{\e^n}}. \eqno(2.12)
$$
 The diagram-by-diagram subtraction method is of course equivalent to
the standard method of constructing counterterms at each loop order from the
lower-order corrections in Eq. (2.12), but it obviates the need to consider
also the wave-function renormalisation of the quantum fields\Ref\howe{P. S.
Howe, G. Papadopoulos and K. S. Stelle, \npb296 (1987) 26}.
Corresponding to Eq. (2.12), the correction to the K\"ahler metric is given
by
$$
g^B_{p\bq}=g_{p\bq}
+\sum_{L=1}^{\infty}\sum_{n=1}^L{1\over{(4\pi)^L}}{K^{(n,L)}_{p\bq}
\over{\e^n}} \eqno(2.13)
$$
and the $\b$-function is then given by
$$\b_{p\bq}=\sum_{L=1}^{\infty}LK^{(1,L)}_{p\bq}=[\b^K]_{p\bq} \eqno(2.14)
$$
where the $\b$-function for $K$ is given by
$$
\b^K=\sum_{L=1}^{\infty}LK^{(1,L)}.  \eqno(2.15)
$$
\vskip 1in
\line {\bigbf 3. Feynman diagram calculations up to five loops \hfil}
\vskip 10pt
It is straightforward to show\refmark\gvz\
 from supergraph power counting that the
divergent counterterms will not involve any superspace derivative $D$ or
$\bar D$ acting on the background fields. Hence the $D$-algebra can be
performed by integrating by parts the $D$, $\bar D$ only on the internal
quantum lines. We can then discard from the expansion Eq. (2.10) all terms
containing only quantum $\P$'s or only $\bP$'s.
The one-loop counterterm is given by
$$
K^{(1,1)}=-2{\rm tr\, ln}\, K_{p\bq}  \eqno(3.1)
$$
which leads to the well-known one-loop $\b$-function
$$
\b^{(1)}_{p\bq}=2R_{p\bq}. \eqno(3.2)
$$
(The Ricci tensor $R_{p\bq}$ is given by $R_{p\bq}=-g^{r\bs}R_{p\bs r\bq}$).
The two and three loop simple pole counterterms $K^{(1,2)}$ and $K^{(1,3)}$
are zero in minimal subtraction, leading to vanishing $\b$-function at two
and three loops; however Grisaru, van de Ven and Zanon\refmark\gvz
 showed that $K^{(1,4)}$
is non-zero and is given by
$$
K^{(1,4)}={\z(3)\over3}(R_p{}^q{}_r{}^sR^u{}_q{}^v{}_sR_u{}^p{}_v{}^r
+R_p{}^q{}_r{}^sR_s{}^r{}_v{}^uR_u{}^v{}_q{}^p)  \eqno(3.3)
$$
which implies a non-vanishing contribution to the $\b$-function for the
supersymmetric K\"ahler $\s$-model at four loops.
Subsequently,
 Grisaru, Kazakov and Zanon\refmark\gkz computed the simple pole contribution
at five loops, i.e. $K^{(1,5)}$. They found a non-vanishing contribution
within minimal subtraction, given by
$$\eqalignno{
K^{(1,5)}&=-{3\z(4)\over{10}}(R_p{}^q{}_r{}^sR_s{}^r{}_u{}^vR_v{}^u{}_w{}^x
R_x{}^w{}_q{}^p-R_p{}^q{}_r{}^sR_u{}^p{}_v{}^rR_w{}^u{}_x{}^v
R_q{}^w{}_s{}^x\cr
&\qquad+R_p{}^q{}_r{}^sR_u{}^p{}_v{}^rR_q{}^u{}_w{}^xR_s{}^v{}_x{}^w
+R_p{}^q{}_r{}^sR_s{}^r{}_u{}^vR_q{}^u{}_w{}^xR_x{}^w{}_v{}^p\cr
&\qquad+\nabla_wR_p{}^q{}_r{}^s\nabla^wR_u{}^p{}_v{}^rR_q{}^u{}_s{}^v
+\nabla^wR_p{}^q{}_r{}^s\nabla_wR_u{}^p{}_v{}^rR_q{}^u{}_s{}^v\cr
&\qquad+2\nabla_wR_p{}^q{}_r{}^s\nabla^wR_s{}^r{}_u{}^vR_v{}^u{}_q{}^p).
&(3.4)
\cr}
$$

They then observed that the resulting contribution to the $\b$-function
could in fact be removed by a local
field redefinition of the metric, equivalent to a change in renormalisation
scheme. The effect of a change $\d g_{p\bq}$ in the metric $g_{p\bq}$ on the
$\b$-function is given by
$$
\d\b_{p\bq}=\b.{\partial\over{\partial g}}\d g_{p\bq}
-\d g.{\partial\over{\partial g}}\b_{p\bq}.\eqno(3.5)$$
Using Eqs. (2.14), (2.15), it is easy to see that if $\d g_{p\bq}$ is
generated according to Eq. (2.3) by a shift $\d K$ in the K\"ahler
potential, the corresponding effect on $\b^K$ in Eq. (2.15) is given by
$$
\d \b^K=\b.{\partial\over{\partial g}}\d K
-\d g.{\partial\over{\partial g}}\b^K. \eqno(3.6)$$
Using
$$
R_{u\bv}.{\partial\over{\partial g_{u\bv}}}R_p{}^q{}_r{}^s
=\nabla^q\nabla_p R_r{}^s
-R_u{}^qR_p{}^u{}_r{}^s  \eqno(3.7)
$$
together with Eqs. (3.1), and (3.2), Grisaru, Kazakov and Zanon\refmark\gkz
 showed that
taking
$$
\d K={3\over4}{\z(4)\over{\z(3)}}K^{(1,4)}  \eqno(3.8)
$$
induced a change in the five-loop $\b$-function given by
$$
\d \b^{K(5)}=-5K^{(1,5)}  \eqno(3.9)
$$
(with $K^{(1,5)}$ as in Eq. (3.4)), and hence the five-loop contribution to
the $\b$-function is removed by the field redefinition Eq. (3.8); in other
words there is a renormalisation scheme in which the $\b$-function is zero
at five loops.
\vskip 1in
\line {\bigbf 4. The six-loop calculation \hfil}
\vskip 10pt
In this section we present details of a six-loop calculation performed with
the aim of investigating whether the six-loop $\b$-function could also be
removed by field redefinitions. In fact we can show by carrying out only a
small fraction of the full six-loop calculation that the six-loop
$\b$-function cannot be eradicated. It is sufficient to focus attention on
diagrams with the topology shown in Fig. 1. The reason for selecting these
particular diagrams is that they are the only ones with three or fewer
vertices which contribute to the $\b$-function. All other diagrams with three
or fewer vertices can easily be reduced using $D$-algebra to standard
Feynman diagrams containing tadpoles, which do not contribute to the
$\b$-function in minimal subtraction. Hence these diagrams will
turn out to determine
all terms in $\b^{K(6)}$ with three or fewer Riemann tensors.
It is straightforward to show using $D$-algebra that any superspace
diagram with the topologies shown in Fig. 1 must contain at least two
$\P$ quantum lines and at least two $\bP$ quantum lines at each vertex,
otherwise it can be reduced to a diagram with tadpoles. Hence the only
superspace
diagrams with the topologies of Fig. 1 which contribute to the $\b$-function
are those shown in Fig. 2. Using $D$-algebra, each of the superspace
diagrams Figs. 2(a)-(e) can be reduced to the momentum integral Fig. 3(a)
and each of Figs. 2(f)-(l) can be reduced to Fig. 3(b). The evaluation of the
momentum integrals is tedious but straightforward. As mentioned earlier, we
subtract from each six-loop diagram the lower order diagrams with
counterterm insertions corresponding to divergent subdiagrams of the
six-loop diagram. We regulate infra-red divergences by replacing potentially
infra-red divergent propagators $1\over{k^2}$ by\refmark\gkz\Ref\chet{K. G.
Chetyrkin and F. V. Tkachov, \plb114 (1982) 133; V. A. Smirnov and K. G.
Chetyrkin, Theor. Math. Phys. 63 (1985) 462}
$$
{1\over{k^2}}+{2\over{\e}}\d(k).   \eqno(4.1)
$$
This avoids the necessity for massive propagators, thereby simplifying the
calculation enormously. Denoting by $G_a$, $G_b$ the momentum integrals
corresponding to Figs. 3(a), (b), (together with their subtraction diagrams),
we find
$$\eqalignno{
G_a&=-{4\over5}{1\over{(4\pi\e)^6}}({496\over3}+40\z(3)\e^3-15\z(4)\e^4
-7\z(5)\e^5)&(4.2a)\cr
G_b&=-{2\over15}{1\over{(4\pi\e)^6}}(8-4\z(3)\e^3+3\z(4)\e^4+3\z(5)\e^5).
&(4.2b)\cr}
$$
The resulting contributions to $K^{(1,6)}$ arising from the diagrams in Fig. 2
will be non-covariant, consisting of products of derivatives of $K$ contracted
together. However it can be proved using $N=2$
supersymmetry\refmark\gvz that the final
complete result for $K^{(1,6)}$ should be covariant. This implies that
 contributions from graphs with more than three vertices will appear in such a
way as to ``covariantise'' the contributions of Fig. 2. Since there are no
graphs with only two vertices contributing to $K^{(1,6)}$, it is clear from
Eq. (2.5) that every term in $K^{(1,6)}$ must contain at least three
Riemann tensors. Hence every vertex in the diagrams of Fig. 2 must
correspond to the linear term on the right-hand side of Eq. (2.5) (or
derivatives of it); graphs
with additional vertices must supply the quadratic terms on the right-hand
side of Eq. (2.5) so as to reconstitute the Riemann tensor. It follows that we
can uniquely reconstruct the terms cubic in the Riemann tensor in the final
covariant result by substituting in the contribution from Fig. 2\refmark\gkz:
$$\eqalignno{
K_{pq\bp\bq}&\rightarrow R_{p\bp q\bq}\cr
K_{pqr\bp\bq}&\rightarrow \nabla_rR_{p\bp q\bq}\cr
K_{pqr\bp\bq\br}&\rightarrow \nabla_r\nabla_{\br}R_{p\bp q\bq}\cr
K_{pqrs\bp\bq}&\rightarrow \nabla_r\nabla_sR_{p\bp q\bq}. &(4.3)\cr}
$$
Combining symmetry factors and $D$-algebra factors for the graphs in Fig. 2
with the results for the momentum integrals in Eq. (4.2), and then
reconstituting the covariant expression {\it via} the substitutions
Eq. (4.3), we obtain
$$\eqalignno{
K^{(1,6)}&=-{7\over30}\z(5)(
\nabla_x\nabla_wR_p{}^q{}_r{}^s\nabla^x\nabla^wR_u{}^p{}_v{}^rR_q{}^u{}_s{}^v
+2\nabla^x\nabla_wR_p{}^q{}_r{}^s\nabla_x\nabla^wR_u{}^p{}_v{}^rR_q{}^u{}_s{}^v
\cr &\quad
+\nabla^x\nabla^wR_p{}^q{}_r{}^s\nabla_x\nabla_wR_u{}^p{}_v{}^rR_q{}^u{}_s{}^v
+2\nabla_x\nabla_wR_p{}^q{}_r{}^s\nabla^x\nabla^wR_s{}^r{}_u{}^vR_v{}^u{}_q{}^p
\cr &\quad
+2\nabla^x\nabla_wR_p{}^q{}_r{}^s\nabla_x\nabla^wR_s{}^r{}_u{}^vR_v{}^u{}_q{}^p
)\cr&\quad
+{1\over5}\z(5)(
\nabla_wR_p{}^q{}_r{}^s\nabla_x\nabla^wR_u{}^p{}_v{}^r\nabla^xR_q{}^u{}_s{}^v
+\nabla_wR_p{}^q{}_r{}^s\nabla^x\nabla^wR_u{}^p{}_v{}^r\nabla_xR_q{}^u{}_s{}^v
\cr &\quad
+\nabla^wR_p{}^q{}_r{}^s\nabla_x\nabla_wR_u{}^p{}_v{}^r\nabla^xR_q{}^u{}_s{}^v
+\nabla^wR_p{}^q{}_r{}^s\nabla^x\nabla_wR_u{}^p{}_v{}^r\nabla_xR_q{}^u{}_s{}^v
\cr &\quad
+\nabla_wR_p{}^q{}_r{}^s\nabla^x\nabla^wR_s{}^r{}_u{}^v\nabla_xR_v{}^u{}_q{}^p
+2\nabla_wR_p{}^q{}_r{}^s\nabla_x\nabla^wR_s{}^r{}_u{}^v\nabla^xR_v{}^u{}_q{}^p
\cr &\quad
+\nabla^wR_p{}^q{}_r{}^s\nabla_x\nabla_wR_s{}^r{}_u{}^v\nabla^xR_v{}^u{}_q{}^p
)+\ldots&(4.4)\cr}
$$
where the ellipsis represents terms with more than three Riemann tensors.
We now need to consider the effects of field redefinitions. A  five-loop field
redefinition
$$
\d g^{(5)}_{p\bq}=\partial_p\partial_{\bq}\d K^{(5)} \eqno(4.5)
$$
produces a change in $\b^{K(6)}$ given according to Eqs. (3.1), (3.2)
and (3.6)
by
$$
\d \b^{K(6)}={\cal O}\d K^{(5)} \eqno(4.6)
$$
where
$$
{\cal O}=\nabla_u\nabla^u
+R_{u\bv}.{\partial\over{\partial g_{u\bv}}}. \eqno(4.7)
$$
Using the identity
$$\eqalignno{
\nabla_u\nabla^u
R_p{}^q{}_r{}^s&=R_u{}^qR_p{}^u{}_r{}^s-\nabla^q\nabla_pR_r{}^s \cr
&\qquad +R_p{}^v{}_u{}^qR_v{}^u{}_r{}^s+R_r{}^v{}_u{}^qR_p{}^u{}_v{}^s
-R_v{}^s{}_u{}^qR_p{}^u{}_r{}^v &(4.8)\cr}
$$
(which follows from the Bianchi identity), together with Eq. (3.7), we have
$$\eqalignno{
{\cal O} R_p{}^q{}_r{}^s&=
R_p{}^v{}_u{}^qR_v{}^u{}_r{}^s+R_r{}^v{}_u{}^qR_p{}^u{}_v{}^s
-R_v{}^s{}_u{}^qR_p{}^u{}_r{}^v &(4.9a)\cr
{\cal O} \nabla^wR_p{}^q{}_r{}^s&=\nabla^w(
R_p{}^v{}_u{}^qR_v{}^u{}_r{}^s+R_r{}^v{}_u{}^qR_p{}^u{}_v{}^s
-R_v{}^s{}_u{}^qR_p{}^u{}_r{}^v)\cr
&\quad +R_p{}^v{}_u{}^w\nabla^uR_v{}^q{}_r{}^s
-R_v{}^q{}_u{}^w\nabla^uR_p{}^v{}_r{}^s+R_r{}^v{}_u{}^w\nabla^uR_p{}^q{}_v{}^s
\cr &\quad
 -R_v{}^s{}_u{}^w\nabla^uR_p{}^q{}_r{}^v &(4.9b)\cr
{\cal O}\nabla_x\nabla^wR_p{}^q{}_r{}^s&=\nabla_x\nabla^w(
R_p{}^v{}_u{}^qR_v{}^u{}_r{}^s+R_r{}^v{}_u{}^qR_p{}^u{}_v{}^s
-R_v{}^s{}_u{}^qR_p{}^u{}_r{}^v)\cr
&\quad+\nabla_x(
 R_p{}^v{}_u{}^w\nabla^uR_v{}^q{}_r{}^s
-R_v{}^q{}_u{}^w\nabla^uR_p{}^v{}_r{}^s+R_r{}^v{}_u{}^w\nabla^uR_p{}^q{}_v{}^s
\cr &\quad
 -R_v{}^s{}_u{}^w\nabla^uR_p{}^q{}_r{}^v)\cr
&\quad +(R_v{}^w{}_x{}^u\nabla_u\nabla^vR_p{}^q{}_r{}^s
-R_p{}^v{}_x{}^u\nabla_u\nabla^wR_v{}^q{}_r{}^s
+R_v{}^q{}_x{}^u\nabla_u\nabla^wR_p{}^v{}_r{}^s
\cr &\quad -R_r{}^v{}_x{}^u\nabla_u\nabla^wR_p{}^q{}_v{}^s
+R_v{}^s{}_x{}^u\nabla_u\nabla^wR_p{}^q{}_r{}^v&(4.9c)\cr}
$$
where ${\cal O}$ is defined in Eq. (4.7). One point to notice is that all terms
involving the Ricci tensor have cancelled on the right-hand sides of
Eq. (4.9). This is a useful property since $K^{(1,6)}$ does not contain the
Ricci tensor. (In fact the Ricci tensor never appears in the simple pole
counterterms when using minimal subtraction, since it corresponds to tadpole
diagrams.)
In order to correspond to a change in renormalisation prescription,
$\d K^{(5)}$ must be a local quantity constructed from covariant quantities,
namely the Riemann tensor and its covariant derivatives. It is convenient
to represent both $\d K^{(5)}$ and the resultant $\d \b^{K(6)}$
diagrammatically, with each Riemann tensor (or its covariant derivative)
denoted by a vertex with a number of legs corresponding to the number of
indices, and contractions represented by lines joining the vertices. Of course
the leading covariant term obtained by the substitution Eq. (4.3) in the
counterterm for a given diagram is then represented according to this
prescription by the original diagram.
When evaluating the contribution to $\d \b^{K(6)}$ from any term in
$\d K^{(5)}$, constructed from a string of Riemann tensors and their
derivatives contracted together, we see from Eqs. (4.6), (4.7) that there are
two possibilities:
\Item{(1.)} The two derivatives in $\nabla_u\nabla^u$
 act on different Riemann tensors
(or derivatives of Riemann tensors). The diagram for the term thus obtained
 is constructed from the original diagram for the term in $\d K^{(5)}$ by
adding an extra line between two vertices.
\Item{(2.)} Both derivatives in $\nabla_u\nabla^u$ act on
 the same Riemann tensor (or
derivative of a Riemann tensor), in which case the contribution to
$\d \b^{K(6)}$ is easily obtained using Eq. (4.9). All terms obtained in this
way are represented by diagrams constructed from the diagram for the term in
$\d K^{(5)}$ by ``opening out'' one of its vertices. By this we mean
replacing a vertex with $m$ legs with two vertices joined by two lines, such
that the total number of free legs is still $m$ and with each new vertex
having at least two free legs.

These two possibilities are depicted in Fig. 4.
For the moment we are only concerned with the topology of the diagrams and we
ignore the orientation of the propagators. Let us now consider how we might
construct a $\d K^{(5)}$ which would produce a $\d \b^{K(6)}$ with the
possibility of cancelling the original $\b^{K(6)}$ derived from Eq. (4.4).
A moment's thought shows that the only possible covariant terms in
$\d K^{(5)}$ which could produce terms of the topology depicted in Fig. 1
according to the two rules given above, and
which could thus have a chance of cancelling the terms
in $\b^{K(6)}$ given explicitly in Eq. (4.4), have the topology shown in
Fig. 5. However, Fig. 5(a) also produces contributions to $\d \b^{K(6)}$ with
the topology of Fig. 6(a), and Fig. 5(b) also produces contributions with the
topology Fig. 6(b). The $D$-algebra for diagrams of the type shown in Fig. 6
always leads to momentum space integrals which reduce to tadpoles; hence
the original $\b^{K(6)}$ contains no terms of the topology Fig. 6(a), (b).
Moreover there is no other possible term in $\d K^{(5)}$ which produces the
topologies of Fig. 6. Hence if we aim to cancel $\b^{K(6)}$ we must omit terms
with the topology of Figs. 5(a), (b) from $\d K^{(5)}$. This means that the
terms of topology Fig. 1 in $\b^{K(6)}$ must be cancelled solely by a
$\d K^{(5)}$ generated by terms of the topology Fig. 5(c). However, any term
in $\d K^{(5)}$ represented by Fig. 5(c) is of the form
$$
a\nabla_wR_p{}^q{}_r{}^s\nabla^wR_u{}^p{}_v{}^rR_q{}^u{}_s{}^v
+b\nabla^wR_p{}^q{}_r{}^s\nabla_wR_u{}^p{}_v{}^rR_q{}^u{}_s{}^v
+c\nabla_wR_p{}^q{}_r{}^s\nabla^wR_s{}^r{}_u{}^vR_v{}^u{}_q{}^p \eqno(4.10)
$$
and generates a contribution to $\d \b^{K(6)}$ given by
$$\eqalignno{
a(
\nabla^x\nabla_wR_p{}^q{}_r{}^s\nabla_x\nabla^wR_u{}^p{}_v{}^rR_q{}^u{}_s{}^v
&+\nabla^x\nabla^wR_p{}^q{}_r{}^s\nabla_x\nabla_wR_u{}^p{}_v{}^rR_q{}^u{}_s{}^v
+2\nabla_wR_p{}^q{}_r{}^s\nabla_x\nabla^wR_u{}^p{}_v{}^r\nabla^xR_q{}^u{}_s{}^v
\cr
+\nabla_wR_p{}^q{}_r{}^s\nabla^x\nabla^wR_u{}^p{}_v{}^r\nabla_xR_q{}^u{}_s{}^v
&+\nabla^wR_p{}^q{}_r{}^s\nabla_x\nabla_wR_u{}^p{}_v{}^r\nabla^xR_q{}^u{}_s{}^v
)+b(
\nabla_x\nabla_wR_p{}^q{}_r{}^s\nabla^x\nabla^wR_u{}^p{}_v{}^rR_q{}^u{}_s{}^v
\cr
+\nabla^x\nabla_wR_p{}^q{}_r{}^s\nabla_x\nabla^wR_u{}^p{}_v{}^rR_q{}^u{}_s{}^v
&+\nabla_wR_p{}^q{}_r{}^s\nabla^x\nabla^wR_u{}^p{}_v{}^r\nabla_xR_q{}^u{}_s{}^v
+\nabla^wR_p{}^q{}_r{}^s\nabla_x\nabla_wR_u{}^p{}_v{}^r\nabla^xR_q{}^u{}_s{}^v
\cr
+2\nabla^wR_p{}^q{}_r{}^s\nabla^x\nabla_wR_u{}^p{}_v{}^r\nabla_xR_q{}^u{}_s{}^v
)&
+c(
\nabla_x\nabla_wR_p{}^q{}_r{}^s\nabla^x\nabla^wR_s{}^r{}_u{}^vR_v{}^u{}_q{}^p
+\nabla^x\nabla_wR_p{}^q{}_r{}^s\nabla_x\nabla^wR_s{}^r{}_u{}^vR_v{}^u{}_q{}^p
\cr
+\nabla_wR_p{}^q{}_r{}^s\nabla^x\nabla^wR_s{}^r{}_u{}^v\nabla_xR_v{}^u{}_q{}^p
&
+2\nabla_wR_p{}^q{}_r{}^s\nabla_x\nabla^wR_s{}^r{}_u{}^v\nabla^xR_v{}^u{}_q{}^p
\cr
+\nabla^wR_p{}^q{}_r{}^s\nabla_x\nabla_wR_s{}^r{}_u{}^v\nabla^xR_v{}^u{}_q{}^p
). &\qquad&(4.11)\cr}
$$
It is immediately apparent that there is no choice of $a$, $b$ and $c$ which
could cancel the terms in
$\b^{K(6)}$ given by Eq. (4.4). Hence $\b^{K(6)}$ canot be removed by any
covariant field redefinition.
\vfill
\eject
\line {\bigbf 5. Conclusions\hfil}
\vskip 10pt
We have shown that there is a well-defined set of terms in $\b^{K(6)}$ (namely
those containing only three Riemann tensors) which are non-zero and
moreover cannot be removed by field redefinitions: so there is no
renormalisation scheme in which the $\b$-function for the K\"ahler potential
in the K\"ahler $\s$-model vanishes at six loops. It follows that
the $\b$-function for
the metric, calculated according to Eq. (2.13), is also non-zero irrespective
of renormalisation scheme at six loops. Now in general, the metric
$\b$-function is ambiguous up to diffeomorphisms (or in other words
co-ordinate changes on the two-dimensional worldsheet) given by
$$
\b_{ij}\rightarrow\b_{ij}+\nabla_{(i}v_{j)} \eqno(5.1)
$$
for some vector $v$ (where $i$, $j$ are {\it real} indices). One might
conceivably entertain the hope that some combination of field redefinition and
diffeomorphism might result in a vanishing $\b$-function. Aside from the
implausibility of this scenario, in any case the important quantity to
 consider from the point of view of string theory is not the metric
$\b$-function itself but rather $B_{ij}$ defined by\refmark\gms\
$$
B_{ij}=\b_{ij}+\nabla_{(i}S_{j)}      \eqno(5.2)
$$
where $S$ is a well-defined, calculable vector quantity. It is the vanishing
of $B_{ij}$ which is the condition for conformal invariance, and $B_{ij}$ is
invariant under diffeomorphisms, since under Eq. (5.1), we also have
$$
S_i\rightarrow S_i-v_i. \eqno(5.3)
$$
Now for the supersymmetric K\"ahler $\s$-model, it is known that $S$ is zero
to all orders
when calculated in the usual complex co-ordinates in which the metric
$\b$-function is given by Eq. (2.14)\Ref\jj{I. Jack and D. R. T. Jones,
\plb220 (1989) 176}.
 Hence, from Eq. (5.2), we see that
$B_{ij}$ is non-vanishing in every renormalisation scheme at six loops, which
is a co-ordinate-independent (or diffeomorphism-invariant) statement.
If it had turned out that there was a scheme in which $B_{ij}$ vanished at
six loops, then the non-vanishing $B_{ij}$ in any other scheme would in some
sense have been an artefact--the six-loop divergence would have been
 generated by the lower-order
divergences. As it is, we conclude that in fact there is a new and independent
contribution to the $\b$-function at six loops.
The fact that this did not occur at five loops remains mysterious.
\vskip 1in
\line {\bigbf Acknowledgements}
\vskip 10pt
I. J. and J. P. thank the S. E. R. C. for financial support.
\refout
\eject
\line {\bigbf Figure Captions \hfil}
\vskip 10pt
\Item {Fig. 1} Topology of 3-point diagrams contributing to the $\b$-function.
\Item {Fig. 2} Superspace diagrams contributing to the $\b$-function (arrows
pointing towards $\P$ lines at vertices).
\Item {Fig. 3} Six-loop momentum integrals (pairs of similar arrows
representing contracted momenta in the numerator).
\Item {Fig. 4} Examples of operations generating a contribution to
 $\d\b^{K(6)}$ from a term in $\d K^{(5)}$.
\Item {Fig. 5} Diagrams representing possible covariant terms in $\d K^{(5)}$
which would produce in $\d\b^{K(6)}$ 3-Riemann terms like those
 in $\b^{K(6)}$.
\Item {Fig. 6} Diagrams representing additional contributions to
$\d\b^{K(6)}$, not already in $\b^{K(6)}$,
produced by terms in $\d K^{(5)}$ shown in Figs 5(a), (b).
\vfill
\eject
\end